\documentclass[aip,jcp,amsmath,amssymb,reprint]{revtex4-1}

\usepackage{graphicx}% Include figure files
\usepackage{dcolumn}% Align table columns on decimal point

\usepackage{bm} % bold math
\usepackage{mathptmx} % times new romans font for text and math

\usepackage{xcolor}
\usepackage[version=4]{mhchem}
\usepackage{booktabs}
\usepackage{array}

\begin{document}

%%%%%%%%%%%%%%%%%%%%%%%%%%%%%%%%%%%%%%%%%%%%%%%%%%%%%%%%%%%%%%%%%%%%%%%%%%%%%%%%%%%%%%%%%%

\title{Ultrafast Photochemistry and Electron Diffraction for Cyclobutanone in the \ce{S2} State:
Surface Hopping with Time-Dependent Density Functional Theory}

%%%%%%%%%%%%%%%%%%%%%%%%%%%%%%%%%%%%%%%%%%%%%%%%%%%%%%%%%%%%%%%%%%%%%%%%%%%%%%%%%%%%%%%%%%

\author{Ericka Roy Miller}
\thanks{These authors contributed equally to this work}
\affiliation{Department of Chemistry, Case Western Reserve University \\ 10900 Euclid Ave, Cleveland, OH 44106, USA}
% ORCID 0000-0003-2073-9446

\author{Sean J. Hoehn}
\thanks{These authors contributed equally to this work}
\affiliation{Department of Chemistry, Case Western Reserve University \\ 10900 Euclid Ave, Cleveland, OH 44106, USA}
% ORCID 0000-0002-8282-7807

\author{Abhijith Kumar}
\affiliation{Department of Chemistry, Case Western Reserve University \\ 10900 Euclid Ave, Cleveland, OH 44106, USA}
% ORCID 0000-0002-1222-9991

\author{Dehua Jiang}
\affiliation{Department of Chemistry, Case Western Reserve University \\ 10900 Euclid Ave, Cleveland, OH 44106, USA}
% ORCID 0000-0002-3836-7639

\author{Shane M. Parker}
\email{shane.parker@case.edu}
\affiliation{Department of Chemistry, Case Western Reserve University \\ 10900 Euclid Ave, Cleveland, OH 44106, USA}
% ORCID 0000-0002-1110-3393

%%%%%%%%%%%%%%%%%%%%%%%%%%%%%%%%%%%%%%%%%%%%%%%%%%%%%%%%%%%%%%%%%%%%%%%%%%%%%%%%%%%%%%%%%%

\begin{abstract}
  We simulate the photodynamics of gas-phase cyclobutanone excited to the \ce{S2} state using
  fewest switches surface hopping (FSSH) dynamics powered by time-dependent density functional 
  theory (TDDFT). We predict a total C3+C2 photoproduct yield of 9\%, with a C3:C2 product ratio
  of 1:8. Two primary \ce{S2}$\rightarrow$\ce{S1} conical intersections are identified: 
  $\beta$ stretch and CCH bend, with the higher energy $\beta$ stretch being associated with
  sub-picosecond \ce{S2} decay. Excited state lifetimes computed with respect to electronic 
  state populations were found to be 7.0 ps (\ce{S2}$\rightarrow$\ce{S1}) and 550 fs 
  (\ce{S1}$\rightarrow$\ce{S0}). We also generate time-resolved difference pair distribution 
  functions ($\Delta$PDFs) from our TDDFT-FSSH dynamics results in order to generate direct 
  comparisons to ultrafast electron diffraction experiment observables. 
  Global and target analysis of time-resolved $\Delta$PDFs produced a distinct set of lifetimes:
  i) a 0.462 ps decay, and ii) a 16.8 ps decay that both resemble the \ce{S2}
  minimum, as well as iii) a long ($>$ nanosecond) decay that resembles 
  the \ce{S1} minimum geometry and the fully separated C3/C2 products. Finally, we contextualize our
  results by considering the impact of the most likely sources of significant errors.
\end{abstract}

%%%%%%%%%%%%%%%%%%%%%%%%%%%%%%%%%%%%%%%%%%%%%%%%%%%%%%%%%%%%%%%%%%%%%%%%%%%%%%%%%%%%%%%%%%

\maketitle

%%%%%%%%%%%%%%%%%%%%%%%%%%%%%%%%%%%%%%%%%%%%%%%%%%%%%%%%%%%%%%%%%%%%%%%%%%%%%%%%%%%%%%%%%%

\section{Introduction}

Photochemical processes involve tightly coupled electronic and nuclear motions occurring on
ultrafast timescales. Unraveling the detailed mechanisms of these processes
has relied on an equally tight synergy between
advanced experimental and computational techniques.
The most commonly applied experimental techniques with
ultrafast time resolution are transient spectroscopy experiments
such as transient UV-vis absorption spectroscopy,\cite{Berera2009PRa},
time-resolved infrared spectroscopy,\cite{Mezzetti2017PR}
or time-resolved X-ray absorption spectroscopy.\cite{Geneaux2019PTRSMPES}
Transient spectroscopies uncover photochemical reactivities by probing the evolution of
the electronic or vibronic structure over time. However, these approaches can only provide
indirect evidence about chemical structure, such as interatomic distances.
Instead, structural information is typically inferred by comparing
with simulations.
Nonadiabatic molecular dynamics (NAMD) simulations have emerged as a powerful
complement to ultrafast experiments for photochemical
reactions.\cite{Tapavicza2013PCCP,Curchod2018CR,Nelson2020CR}
At a conceptual level, NAMD simulations produce molecular
movies that show the movement of electrons and nuclei in response to absorbed light.
Hence, NAMD simulations can provide the atomic-scale insights
unavailable from transient spectroscopies,
provided the simulations are accurate analogues of the experiment.
One consequence of the lack of direct experimental structural information
is that the accuracy of NAMD simulations is difficult to assess
without a direct comparison to experiment.

Over the last decade, ultrafast electron diffraction (UED)
experiments using megaelectron-volt (MeV) electrons
have proven capable of providing time-resolved structural
information with femtosecond time resolution and sub-Ångstrom spatial
resolution.\cite{yang2016diffractive,shen2019femtosecond} 
Recent applications of UED have thus provided some of the first
direct measurements of both electronic and nuclear dynamics
in photochemical
reactions.\cite{yang2018imaging,liu2020spectroscopic,UEDelec2020Sci,centurion2022ultrafast,filippetto2022ultrafast,champenois2023femtosecond}

The time-resolved structural information provided by UED experiments
provides the first opportunity to test the structural predictions of
NAMD simulations.
Indeed, this Special Topics collection was prepared in recognition
of this opportunity, and to encourage the further development of
accurate NAMD methods.
The challenge as presented by the organizers of this Special Topics
collection is to simulate the photochemical dynamics of cyclobutanone (CB)
after excitation with a 200 nm laser pulse, including
the prediction of time-resolved structural measurements.
Simultaneously, a set of UED experiments will be performed to
serve as a time-limited double-blind test of the NAMD simulations.\cite{weathersby2015mega}

Cyclobutanone is a valuable synthetic precursor,
\cite{Tanino2011NatChem,Hong2023NatCommun}
with a diverse set of photo-dissociation pathways that have been intensively
investigated both experimentally and computationally.
The primary focus of both experimental and computational studies has been on
photochemical pathways present after initial excitation to the weakly absorbing
\ce{S1} state. Two categories of photo-products are seen for cyclobutanone: the C3
products, consisting of \ce{C3H6} and \ce{CO}, and the C2 products, consisting of
\ce{C2H4} and \ce{CH2CO}. The ratio of C3:C2 products has been shown to be
wavelength dependent: longer wavelength excitations favor C3 products,
while shorter wavelength excitations tend to favor C2 products.\cite{TangLee1976JPC}
There has been much debate as to the mechanisms underlying these product ratios.
One hypothesis has been that an \ce{S1}$\rightarrow$\ce{T1}
intersystem crossing pathway leading to C3 products is in competition with an
\ce{S1}$\rightarrow$\ce{S0} internal conversion process leading to C2
products.\cite{TangLee1976JPC, Lee1971JACS, Lee1971JACS2, Liu2016JCP}
However, ab-initio multiple spawning (AIMS) simulations with complete active space
self-consistent field (CASSCF) found the existence of 3 distinct
\ce{S1}$\rightarrow$\ce{S0} conical intersections,
where progression through each
produced different C3:C2 product ratios.\cite{Liu2016JCP}

Comparatively fewer investigations have focused on the photodynamics
upon excitation at 200 nm, which corresponds to excitation of the \ce{S2}
state.
Time-resolved mass spectrometry
experiments have shown a rapid sub-picosecond decay from \ce{S2} to \ce{S1} involving a
ring-puckering vibrational mode. \cite{S2TRMS2012CPC}
Supporting simulations have shown biexponential decay from \ce{S2} to \ce{S1},
involving an ultrafast 0.95 ps decay and a slower 6.32 ps decay.\cite{Kuhlman2012JCP}

In summary, simulating the photodynamics of CB after excitation with 200 nm
light will require balancing competing pathways of internal conversion and
intersystem crossing, that end in one of two sets of photoproducts, and that may
form intermediates with a diradical ground state.

For this challenge, we have selected fewest switches surface hopping\cite{Tully1990JCP}
(FSSH) dynamics powered by time-dependent density functional theory (TDDFT).\cite{Runge1984PRL}
TDDFT is a ``workhorse'' of photochemical simulations due to its excellent
cost-to-performance ratio. In addition, the growing number of
successful applications of TDDFT to photochemical problems
seems to indicate that it strikes an ideal balance between
static and dynamic correlation, which is crucial for accurate
photochemistry.\cite{Tapavicza2007PRL,Muuronen2017CS,Yue2018PCCP,Parker2019PCCP}
However, TDDFT also has several well known deficiencies.
Perhaps most concerning, TDDFT fails to produce true
conical intersections between the ground state and the first excited state,
instead yielding non-interacting seams with incorrect
dimensionality.\cite{Levine2006MP}
Indeed, TDDFT becomes unstable near degenerate ground states.
In addition, TDDFT has systematic biases resulting in the underestimation of
triplet state energies\cite{Jacquemin2010JCTC,Casanova-Paez2020JCP}
and charge-transfer states.\cite{Dreuw2003JCP}
Finally, DFT with a restricted Kohn-Sham (KS) determinant
typically fails to break bonds homolytically because
the closed-shell determinant can not describe an open-shell
(or diradical) configuration.\cite{Vincent2016JPCL}

We take a pragmatic approach to predicting the photochemical reactivity of
cyclobutanone.
In short, we will perform and analyze photochemistry simulations
using TDDFT-FSSH, and then perform a series of sensitivity analyses to
estimate the potential impact of the above-mentioned deficiencies.
To avoid instabilities surrounding \ce{S1}$\rightarrow$\ce{S0} conical intersections,
we apply an energy gap threshold to force hops to the ground state during our
TDDFT-FSSH dynamics. Intersystem crossing (ISC) effects are excluded entirely 
from our simulations due to technical limitations of our selected TDDFT-FSSH 
implementation. Therefore, we estimate the overall impact on final product quantum 
yields using a Landau-Zener survey of ISC probabilities throughout our dynamics.
And finally, we apply a combination of broken-symmetry and unrestricted TDDFT to
estimate the possible errors in photoproduct yields caused by our use of a restricted
KS determinant in the ground state.

This manuscript is organized as follows. In section \ref{sec:methods}, we describe
the computational methods and implementation details of our simulations.
Section \ref{sec:namd} describes the results of our photochemistry simulations.
Next, in section \ref{sec:sanity} we contextualize the results of our photochemistry simulations with
sensitivity analyses to arrive at final predictions for the photochemistry
of CB.
Finally, we conclude in section \ref{sec:conclusion} with a summary of our approach
and a discussion of the implications of our results both on the photochemistry of
CB as well as on the state of predictive photochemistry.

%%%%%%%%%%%%%%%%%%%%%%%%%%%%%%%%%%%%%%%%%%%%%%%%%%%%%%%%%%%%%%%%%%%%%%%%%%%%%%%%%%%%%%%%%%

\section{Computational Methods} \label{sec:methods}
For all dynamics simulations and for most of our analysis,
we used the \textsc{Turbomole} program suite, version 7.5.\cite{turbomole}
In addition, we used ORCA\cite{Neese2020JCP} version 5.0 for the calculation of
spin-orbit couplings\cite{deSouza2019JCTC}, triplet state energies,
and for minimum energy conical intersection optimizations.
Further details of each stage of our computational methods are given below.

\subsection{Ground State Dynamics}
To sample the equilibrium ensemble and to
generate initial conditions for our NAMD simulations, we performed
10 ps ground state molecular dynamics simulations using \textsc{Turbomole}.
These simulations used the TPSS density functional\cite{Tao2003PRL}
with the def2-SVPD basis set\cite{Rappoport2010JCP} and
D3(BJ)\cite{Grimme2010JCPa,Grimme2011JCC} dispersion corrections.
The resolution-of-the-identity technique was used to accelerate
the Coulomb integrals (RIJ),\cite{Dunlap1979JCP}
grid size 3 with quadrature weight derivatives were used for functional
integration, and a tight SCF convergence threshold of 10$^{-9}$ Eh
(\$scfconv 9) was used to improve energy conservation.
The trajectory was initiated from the ground-state minimum with
initial velocities sampled according to the Botzmann distribution
at 500 K. The dynamics were propagated using \textsc{Turbomole}'s
\verb|frog| module with a 20 a.u. (0.484 fs) time step.
With these settings, the total energy drift during ground state dynamics was
less than 1.6 $\mu$Hartree/ps and oscillations of the total energy were
about 10 $\mu$Hartree.

\subsection{Electronic State Analyses}
Electronic states were characterized by computing
excited states at the ground-state equilibrium structure (i.e., the Franck-Condon structure),
which was optimized with RIJ/TPSS/def2-SVPD/D3(BJ).
To compute excited states, we used Linear Response TDDFT within the Tamm-Dancoff
approximation\cite{Hirata1999CPL} with the PBE0 density functional,\cite{Adamo1999JCP}
and again with the def2-SVPD basis set\cite{Rappoport2010JCP}
and D3(BJ) dispersion corrections.\cite{Grimme2010JCPa,Grimme2011JCC}
All calculations using the PBE0 functional used the resolution-of-the-identity
technique to accelerate both Coulomb and exchange integrals (RIJK).
The size 4 grids with quadrature weight derivatives were again used
for functional integrations, with tight SCF (10$^{-9}$ Eh) and
excited state (residual norm of 10$^{-7}$) convergence thresholds.

\subsection{Nonadiabatic Molecular Dynamics (NAMD)} \label{sec:methods.namd}
All NAMD simulations used \textsc{Turbomole}'s implementation of
Tully's Fewest Switches Surface Hopping (FSSH).\cite{Tully1990JCP}
The dynamics used the RIJK/PBE0/def2-SVPD/D3(BJ) method as described above
to compute all the requisite energies, forces, and couplings.
Nonadiabatic coupling matrix elements were computed as the inner product of
the derivative coupling vectors and the nuclear velocity.
Derivative couplings between the ground-and-excited state are computed
using linear response theory\cite{Send2010JCP} while derivative couplings
between two excited states are computed using quadratic response theory\cite{Parker2019PCCP}
within the pseudowavefunction approximation.\cite{Ou2015JCP}
In both cases, electron translation factors were incorporated by neglecting
terms that violate translational invariance.\cite{Fatehi2011JCP}

Initial conditions were sampled randomly from the ground state MD trajectories
above, and were initiated on the \ce{S2} state, which corresponds
most closely to excitation with 200 nm light.
The dynamics were propagated with the leapfrog Verlet algorithm and a time
step of 20 a.u. (0.484 fs) for a total time of about 10 ps.
After a surface hop, the momenta were rescaled along the direction of
the derivative coupling vector to conserve the total energy.
If there was insufficient momentum in the direction of the derivative coupling
vector to satisfy energy conservation (``frustrated hop''),
then no hop occurred and no further action was taken (i.e., there is no reversal
of momentum).
To avoid instabilities near degenerate ground states, surface hops to the
ground state are forced whenever the first excitation energy to \ce{S1}
falls below 0.5 eV.

All statistical uncertainties are presented as 95\% confidence intervals
that were estimated using bootstrap resampling with 10000
samples, unless stated otherwise.\cite{Efron1979AS,Nangia2004JCP}

\subsection{Conical Intersection and Excited State Minima Optimizations}
Excited state minima for the \ce{S1} and \ce{S2} excited states were computed
using RIJK/PBE0/def2-SVPD/D3(BJ).
Using molecular geometries extracted from surface hopping structures
found during NAMD simulations, we also performed minimal
energy conical intersection (MECI) optimizations with
ORCA, using RIJK/PBE0/def2-SVPD/D3(BJ).

\subsection{Ultrafast Electron Diffraction Patterns}
Ultrafast electron diffraction (UED) patterns were simulated using structural
data from our NAMD simulations and used to estimate effective PDFs.
As our goal is to compare these
PDFs with the experimentally determined PDFs, we used the same
basic strategy as the measurements, in which we first compute
a modified scattering intensity, $sM(s)$, that includes weighted contributions
from all atom pairs in the molecule, and then use the modified scattering
intensity to reconstruct the real space pair distribution function (PDF).
The modified scattering intensity was computed within the
independent atom model, as
\begin{equation}
  sM(s) = s\frac{I_{mol}}{I_{atm}}
\end{equation}
where $s$ is the magnitude of scattering vector, $I_\text{mol}$ is
the molecular scattering intensity, and $I_\text{atm}$ is the atomic
scattering intensity. Explicit expressions for $I_\text{mol}$ and $I_\text{atm}$
are provided in the supplementary information.
We used code developed by Thomas J. A. Wolf to generate $I_\text{mol}$ and $I_\text{atm}$
scattering intensities directly from XYZ structure 
files.\cite{wolfthomas2020diffraction_simulation}

The real space PDF was calculated
from the modified scattering intensity using a Fourier transform,
\begin{equation}
  \text{PDF}(R) = \int_{0}^{s_{max}} sM(s) \sin(sR) e^{-ks^2} \, ds
\end{equation}
where $s_{max}=12 \text{\AA}^{-1}$ is the experimentally determined
maximum amplitude of scattering vector, $R$ is the
interatomic distance, and $k=0.03$ \AA$^2$ is a damping factor chosen
to tame experimental noise at larger values of $s$---which we note
also has the effect of broadening the PDF.
PDFs and modified scattering intensities for the NAMD simulations
were averaged over the swarm of trajectories.
The change in modified scattering intensity and PDFs were computed by subtracting
the intensities and PDFs at $t=0$, i.e., $\Delta\text{PDF}(R,t) = \text{PDF}(R,t) - \text{PDF}(R,0)$
and $\Delta sM(s,t) = sM(s,t) - sM(s,0)$.
Finally, to mimic the 150 fs time resolution of UED experiments, we convolved
the calculated $\Delta$PDF with a Gaussian kernel
having a full-width-at-half-maximum (FWHM) of 150 fs.
Without the convolution, the calculated $\Delta$PDF is
dominated by high frequency oscillations that would be invisible to the experiments.

\subsection{Landau-Zener analysis}
Within the Landau-Zener model, the probability of a molecule undergoing an
\ce{S_n}$\rightarrow$\ce{T_m} transition is given by
\begin{equation}
  P_\text{LZ} = 1 - \exp\left(-\frac{2\pi |V_{nm}^\text{SOC}|^2}{|\dot{\Omega}_{nm}| \hbar}\right)
\end{equation}
where $|V_{nm}^\text{SOC}|$ is the magnitude of the spin-orbit coupling matrix
element and $\dot{\Omega}_{nm}$ is the rate of change of the energy gap,
$\Omega_{nm} = E_n - E_m$,
between the states.\cite{Wittig2005JPCB}
The spin-orbit coupling matrix elements were computed using
ORCA.\cite{deSouza2019JCTC,Neese2020JCP} The rate of change of the energy gap was
computed using a finite difference approximation,
\begin{equation}
  \dot{\Omega}_{nm}(t) = \frac{d}{dt} \Omega_{nm}(t)
      \approx \frac{\Omega_{nm}(t+\Delta t) - \Omega_{nm}(t)}{\Delta t}
\end{equation}
where $t$ is the time of an individual snapshot of the trajectory and $\Delta t$
is the time step. Trajectories were sampled evenly at every 200th time step.
Spin orbit couplings and energy gaps were computed between the five lowest energy
singlet and triplet states at each sampled point.

\subsection{Reaction Path Analysis}
Reaction path analysis was performed using \textsc{Turbomole}.
Structures were generated using linear synchronous transit
interpolation\cite{Halgren1977CPL} via the \textsc{woelfling} module.
Restricted and unrestricted Kohn-Sham singlet ground state energies were computed using
RIJK/PBE0/def2-SVPD/D3(BJ). However, the unrestricted Kohn-Sham energies were obtained
using unrestricted triplet ground state orbitals as the initial guess prior to final
wavefunction optimization.

\subsection{Broken-Symmetry TDDFT}
We use the broken-symmetry TDDFT (BRS-TDDFT)
approach as implemented in \textsc{Turbomole} and reported
previously.\cite{Vincent2016JPCL,Muuronen2017CS}
In brief, at each time step a triplet instability calculation is used to
check whether a lower energy unrestricted Kohn-Sham determinant is
available. If such a state is found, the spin symmetry is broken and
the simulation is continued. Besides this modification, the remainder
of the NAMD simulation is identical to the one described above.

%%%%%%%%%%%%%%%%%%%%%%%%%%%%%%%%%%%%%%%%%%%%%%%%%%%%%%%%%%%%%%%%%%%%%%%%%%%%%%%%%%%%%%%%%%

\section{Photochemistry Simulations} \label{sec:namd}

In this section, we present the results of our PBE0-FSSH simulation of cyclobutanone.
We give a brief overview of the structural and electronic features of cyclobutanone at 
its Franck-Condon geometry.
We then derive conclusions about excited state decay lifetimes and pathways directly from our
simulated data, before rendering our dynamics results in terms of UED observables.

\subsection{The Franck-Condon Regime of Cyclobutanone}

\begin{figure}[htbp]
  \includegraphics[width=\linewidth]{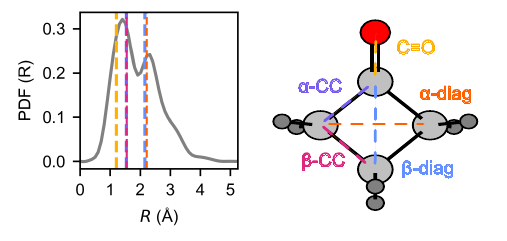}
  \caption{\label{fig:s0pdf}
    Simulated PDF for optimized \ce{S0} minimum geometry of cyclobutanone.
  }
\end{figure}

The PDF of the \ce{S0} equilibrium structure of cyclobutanone is shown in Figure \ref{fig:s0pdf}.
Primary contributions to the first peak include the \ce{CO} bond and the \ce{CC} bonds.
At the Franck-Condon geometry there is near perfect overlap between the
bond length between carbonyl and $\alpha$-carbons and the bond length
between $\alpha$-carbons and $\beta$-carbons.
Going forward, these bonds will be referred to as $\alpha$-\ce{CC}
and $\beta$-\ce{CC} bonds respectively.
The second peak is composed primarily of signals from non-bonding and
cross-ring diagonals, with an extended shoulder originating from
non-bonding \ce{CH} and \ce{OH} interactions.

Using PBE0, we characterized the first three singlet states of cyclobutanone.
The results are collected in Table \ref{tab:vee}.
As shown in Table \ref{tab:vee}, qualitative agreement was found between 
the simulated excitation energies
and the spectroscopically bright state targeted by the 200 nm (6.2 eV) laser probe.
We find that PBE0  overestimates the excitation energies relative
to the experimental values. More specifically, the \ce{S1} excitation computed using
PBE0 is 4.22 eV, compared to the experimental value of 3.92 eV.
Similarly, the \ce{S2} state computed using PBE0 is 6.83 eV, compared to the
experimental value of 6.36 eV.
Nonetheless, we unambiguously identify that the \ce{S2} state is most likely
to be excited at 200 nm, both because it is closest in energy and because it is
brightest.
Thus, we performed PBE0-FSSH simulations starting from the \ce{S2} state
using a swarm of 100 trajectories. In the following, we present the results of
these simulations.

\begin{table}[htbp]
\caption{\label{tab:vee} Computational and experimental characterization of
the three lowest energy singlet states of cyclobutanone}
\begin{ruledtabular}
\begin{tabular}{lccccc}
State &
VEE\footnote{vertical excitation energies using RIJK/PBE0/def2-SVPD/D3(BJ) from \ce{S0} minimum structure} &
\textit{f}\footnote{length representation oscillator strengths} &
Absorption &
\textit{f}\footnote{estimated oscillator strengths
by Whitlock and Duncan\cite{Whitlock1971JCP}} &
Transition \\
  & (eV) &  & Peak (eV) &   &   \\
\hline
\ce{S1} & 4.22 & 6.9$\times 10^{-6}$ & 3.92\footnote{gas phase absorption band maximum\cite{Hemminger1972JCP}}
& 5$\times 10^{-4}$ & n $\rightarrow \pi$* \\
\ce{S2} & 6.83 & 4.0$\times 10^{-2}$ & 6.36\footnote{gas phase absorption band maximum\cite{Udvarhazi1965JCP}}
& 3$\times 10^{-2}$ & n $\rightarrow$ 3s \\
\ce{S3} & 7.44 & 1.8$\times 10^{-5}$ & 7.08 \footnote{gas phase absorption band maximum\cite{Udvarhazi1965JCP}}
& 7$\times 10^{-3}$ & n $\rightarrow$ 3p \\
\end{tabular}
\end{ruledtabular}
\end{table}

\subsection{Excited-state dynamics and lifetimes}

\begin{figure}[htbp]
  \includegraphics[width=\linewidth]{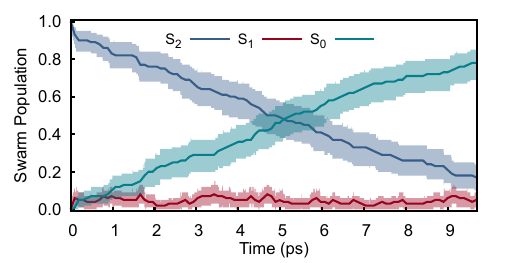}
  \caption{\label{fig:decay}
    Excited state populations measured over trajectory swarms.
    Shaded regions show the 95\% bootstrap confidence interval.
  }
\end{figure}

The qualitative behavior of most of the trajectories followed
a similar pattern. After excitation to the \ce{S2} state,
CB undergoes quick decay to the \ce{S1} state, followed by
even faster decay to the \ce{S0} state. Figure \ref{fig:decay}
shows the excited state populations as a function of time for
the swarm of trajectories over the 10 ps simulation time.
Here, we see a steady decay of the \ce{S2} population, ending
with an \ce{S2} population of 17\%.
We determine a time constant for the decay by first
determining the half-life (i.e., the time at which
$P_2(t)=1/2$), and then assuming
a single exponential decay for the
\ce{S2} swarm population,
$P_2(t) = \exp(-t/\tau_2)$.
We find a half-life of 4.9 ps (CI: 4.2--6.2 ps)
which corresponds to a time constant of
$\tau_2 =$ 7.0 ps (CI: 6.1--8.9 ps).

The \ce{S1} population,
by contrast, has an essentially steady-state behavior, oscillating
around 5\% for most of the simulation time. This
indicates that the \ce{S1} state is much shorter lived than
the \ce{S2} state. We estimate the \ce{S1} lifetime by
computing the average residence time spent in the \ce{S1} state as
$\bar{t}_{1} = t_{1\rightarrow0} - t_{2\rightarrow1}$
where $t_{1\rightarrow0}$ and $t_{2\rightarrow1}$ are the
times at which the \ce{S1}$\rightarrow$\ce{S0} and
\ce{S2}$\rightarrow$\ce{S1} hops occur, respectively, limiting
this analysis to the 77 trajectories that underwent this
pathway. We find a mean residence time of $\bar{t}_1 =$550 fs (CI: 460--650 fs).
If we assume that, once generated, the \ce{S1} state decays exponentially as
$P_1(t') = \exp(-t'/\tau_1)$,
where $t'$ is the time since the \ce{S1} state was first populated,
then the mean residence time is also the lifetime, i.e., $\tau_1 = \bar{t}_1$.

\subsection{Decay Pathways and Conical Intersections}
\begin{figure}[htbp]
  \includegraphics[width=\linewidth]{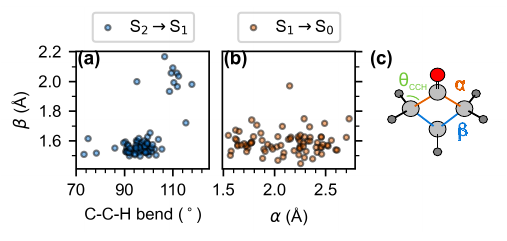}
  \caption{\label{fig:hop_dist}
    Structural characterization of hopping structures observed
    in the NAMD simulations for
    \textbf{(a)} \ce{S2}$\rightarrow$\ce{S1} hops
    and \textbf{(b)} \ce{S1}$\rightarrow$\ce{S0} hops.
  }
\end{figure}

\begin{figure}[htbp]
  \includegraphics[width=\linewidth]{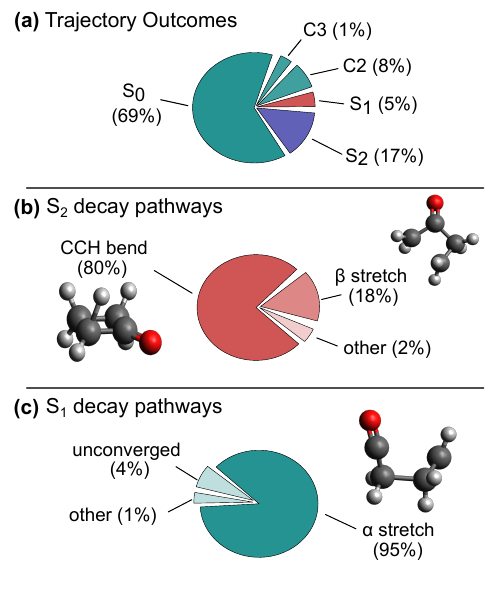}
  \caption{\label{fig:namd_overview}
    \textbf{(a)} Breakdown of PBE0-FSSH dynamics outcomes. \ce{S2} and \ce{S1} categories refer to
    trajectories that remained in these respective excited states, while C2, C3, and \ce{S0}
    categories reflect product ratios of trajectories that decayed to the \ce{S0} state,
    optionally undergoing photodissociation.
    Percentages are given with respect to the 100
    trajectories that completed the full 10 ps simulation time.
    \textbf{(b)} Breakdown of pathways of \ce{S2} decay pathways. Representative structures of the
    $\beta$ stretch and CCH bend \ce{S2}$\rightarrow$\ce{S1} conical intersections
    are shown. Percentages are given with respect to the 83 trajectories that decayed
    from the \ce{S2} state.
    \textbf{(c)} Breakdown of pathways for \ce{S1} decay pathways. A representative structure of the
    $\alpha$ stretch \ce{S1}$\rightarrow$\ce{S0} conical intersection is shown.
    Percentages are given with respect to the 76 trajectories that decayed from the \ce{S1} state.
  }
\end{figure}

Figure \ref{fig:hop_dist} shows the distribution of hopping structures observed
during PBE0-FSSH dynamics. We observe two distinct modes for
\ce{S2}$\rightarrow$\ce{S1} deactivation, which we characterize as
either a $\beta$ stretch, where one of the $\beta$-CC bonds
stretches from an equilibrium value of about 1.6 \AA{} to the range of 2.0-2.2 \AA,
or a CCH bend, in which one of the CCH angles formed between an $\alpha$-CC bond
and a CH bond decreases from an equilibrium value of about 110$^\circ$ to
the range of 90-100 $^\circ$.
MECI optimizations starting from these hopping structures
identified two dominant \ce{S2}$\rightarrow$\ce{S1} structures,
representatives for which are shown in Figure \ref{fig:namd_overview}.
The $\beta$ stretch MECI structure representative has a $\beta$-CC bond length of 2.1 \AA,
while the representative CCH bend MECI structure has a CCH angle of
71$^\circ$.
Furthermore, we found that 80\% of the \ce{S2}$\rightarrow$\ce{S1} deactivations
proceeded through the CCH bend structure, 18\% proceeded through the $\beta$ stretch
structure, and 2\% proceeded through other structures.
Interestingly, although the majority of \ce{S2}$\rightarrow$\ce{S1} deactivations
occurred through the CCH bend structure, those that decayed through the
$\beta$ stretch structure were significantly more reactive.
11 out of the 15 trajectories that decayed via $\beta$ stretch
did so within the first picosecond after photoexcitation, compared
to only 5 out of 66 trajectories that deactivated through a CCH bend structure.
Considering only the 15 trajectories that decayed via $\beta$ stretch,
the mean residence time in the \ce{S2} state was 0.8 ps, in partial agreement
with the fast decay component of 0.95 ps found by Kuhlman et al.\cite{Kuhlman2012JCP}
Notably, the representative $\beta$ stretch structure displayed in Figure \ref{fig:namd_overview} is around 0.8 eV
higher in energy than the representative CCH bend structure.

Comparatively less distinction was found in \ce{S1}$\rightarrow$\ce{S0} hopping structures.
The only significant coordinate we observed was the $\alpha$-CC stretch, which
were found to be broadly distributed from 1.6 \AA{} to 2.5 \AA, compared to an
equilibrium value of 1.5 \AA.
MECI optimizations confirmed only one dominant \ce{S1}$\rightarrow$\ce{S0} structure,
denoted in Figure \ref{fig:namd_overview} as $\alpha$ stretch.

Figure \ref{fig:namd_overview} shows an overview of the observed
decay pathways, including C3/C2 product formation.
Out of the 100 PBE0-FSSH trajectories, only
9\% resulted in photodissociation products within 10 ps,
with a C3:C2 product ratio of 1:8.
A detailed breakdown of trajectory outcomes can be found in
Tables SI and SII.

\subsection{Ultrafast Electron Diffraction (UED) Simulations}
\begin{figure}[htbp]
  \includegraphics[width=\linewidth]{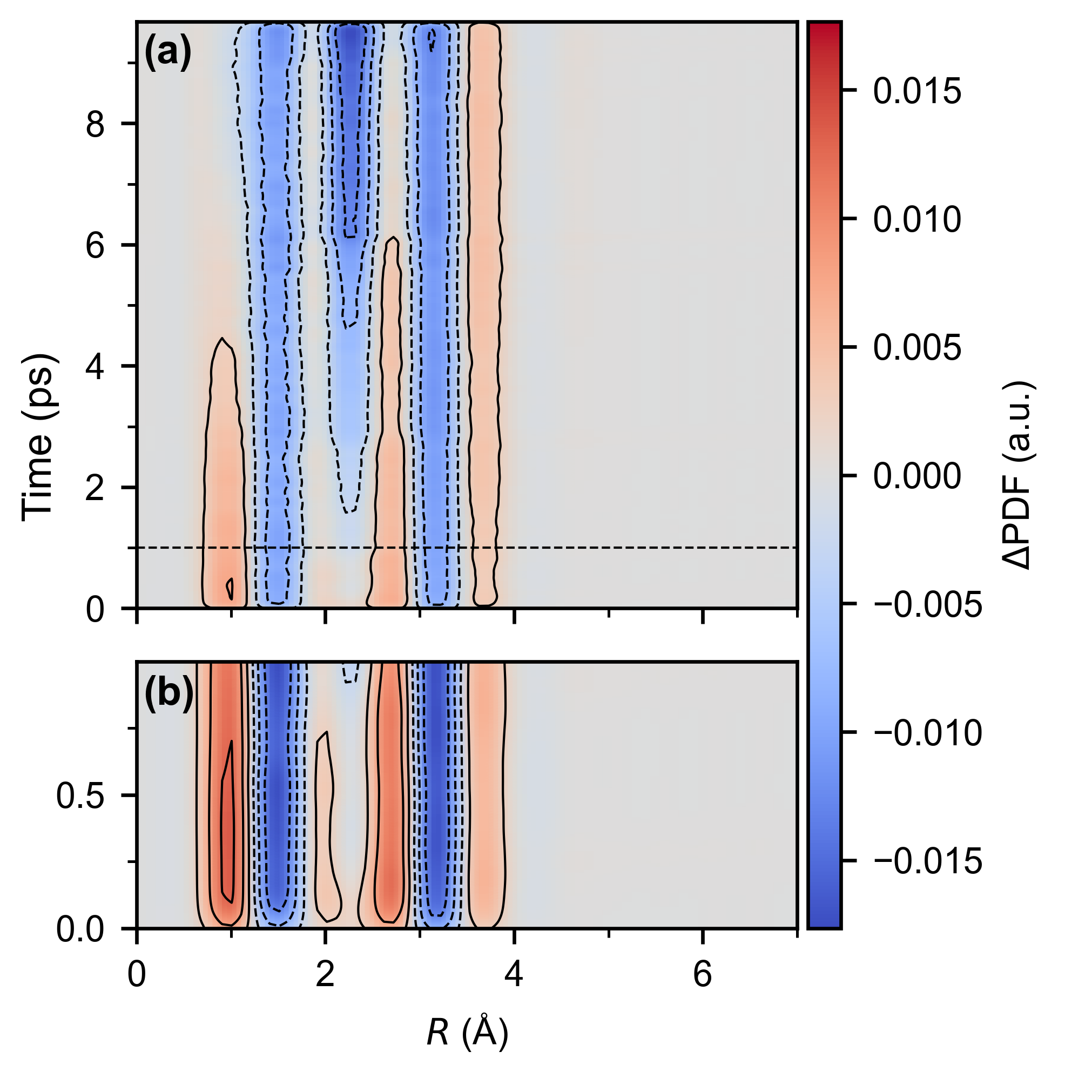}
  \caption{\label{fig:contour_dpdf}
    Contour plot of $\Delta$PDF averaged across all 100 trajectories for 
    \textbf{(a)} the full simulated time regime of 0 - ca. 10 ps, 
    and \textbf{(b)} the ultrafast 0-1 ps regime.
    The horizontal line in \textbf{(a)} shows the time window in \textbf{(b)}.
  }
\end{figure}

\begin{figure}[htbp]
  \includegraphics[width=\linewidth]{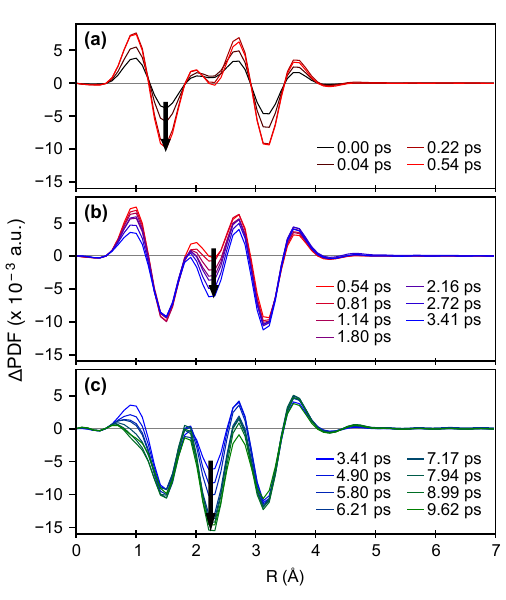}
  \caption{\label{fig:temporal_dpdf}
  Simulated $\Delta$PDF at select time delays after photo-excitation from a) 0-0.5 ps,
  b) 0.5-1.8 ps and c) 3.4-9.6 ps.
  }
\end{figure}

As a direct comparison to experimental observables, we computed
time-resolved $\Delta$PDFs by averaging across all 100 trajectories in
the NAMD swarm. The results are shown in Figure \ref{fig:contour_dpdf}.
In addition, more detailed views of specific time windows are shown in
Figure \ref{fig:temporal_dpdf}.

As shown in Figure \ref{fig:contour_dpdf} and Figure \ref{fig:temporal_dpdf}a,
immediately following UV excitation, positive amplitude in $\Delta$PDF were observed at
distances of ca. 1, 1.9, 2.7, and 3.7 Å.
Simultaneously, within the first approximate 0.5 ps after excitation,
negative $\Delta$PDF bands were observed to grow in magnitude at ca. 1.5 and 3.1 Å.
Between ca. 0.5 ps and 3.5 ps (Figure \ref{fig:temporal_dpdf}b),
the bands at ca. 1, 1.9, and 2.7 Å decrease in intensity,
with a new band growing negatively in amplitude at ca. 2.3 Å.
Interestingly, a slight increase in intensity is observed at ca. 3.7 Å,
while the two negative signals at ca. 1.5 and 3.1 Å remain unchanged.

Lastly, from ca. 3.5 ps to 9.6 ps as shown in Figure \ref{fig:temporal_dpdf}c,
the bands at ca. 1, 2.3, and 2.7 Å continue to grow negatively in magnitude,
with the band at ca. 1 Å shifting to shorter distances and broadening simultaneously.
The remaining bands on the simulated distributions appear to have minimal changes in amplitude.

\begin{figure}[htbp]
  \includegraphics[width=\linewidth]{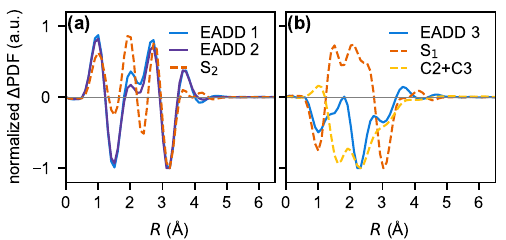}
  \caption{\label{fig:eads_dpdf}
Comparison of extracted EADS from global and target analyses of the
simulated time-resolved $\Delta$PDF with normalized $\Delta$PDF of critical structures.
\textbf{(a)} EADD1 and EADD2 compared to the $\Delta$PDF for the \ce{S2} minimum.
\textbf{(b)} EADD3 compared to the $\Delta$PDF for the \ce{S1} minimum
and the C3 and C2 products.
  }
\end{figure}

In order to contextualize these time-resolved $\Delta$PDFs band structures,
we compare normalized $\Delta$PDFs of
relevant structures and extracted evolution associated decay distributions (EADD)
in Figure \ref{fig:eads_dpdf}.
EADD were obtained from applying global and target
analyses\cite{vanStokkum2004BBA} to the simulated data using a two-component with an offset,
sequential kinetic model via the Glotaran graphical user interface to the R-package TIMP 
software.\cite{Snellenburg2012JSS} 
The EADD are direct analogies of evolution associated decay
spectra commonly used to analyze transient spectra, but we emphasize here that these are
pair distributions, \emph{not spectra}. Lifetimes from this analysis using an instrument response 
function (IRF) of 60 fs, were 0.462$\pm$0.004 ps and 16.8$\pm$0.3 ps.
This finding of a short sub-picosecond lifetime and a longer picosecond lifetime
is consistent with our observation of two distinct decay modes out of the \ce{S2} state,
with the shorter lifetime of about 0.8 ps.
In addition, a long time
offset was found with greater than nanosecond lifetime.
Errors are reported as twice the standard deviation.
We note that the lifetimes computed here, reflecting only structural information,
differ significantly from the earlier lifetimes computed from electronic state information.
This difference strongly emphasizes the need to use the same observables when comparing
simulated and measured results, as there is no strong \textit{a priori} reason why the timescales from
two different observables must agree.

As shown in Figure \ref{fig:eads_dpdf}a, EADD1 and EADD2, which correspond to the
0.462 ps and 16.8 ps lifetimes, line up well with
the $\Delta$PDF of the \ce{S2} minimum.
From EADD3 (Figure \ref{fig:eads_dpdf}b), these signals were found to overlap well
with $\Delta$PDF for the \ce{S1} and fully separated C3/C2 products.

\section{Sensitivity Analyses of Photochemistry Simulations} \label{sec:sanity}
Due to technical limitations, our simulations used a closed shell reference and
only included singlet excited states. 
Therefore, in this section, we attempt to put the PBE0-FSSH simulations in greater context
by performing sensitivity analyses of our most likely sources of error:
intersystem crossing and ground state diradical formation.

\subsection{How likely is intersystem crossing?}
Here, we attempt to quantify
the impact of omitting ISC effects by estimating the probability of ISC using
a Landau-Zener-based approach.
Landau-Zener estimates transition probabilities
between two diabatic states under the assumption that i) the states cross in
energy, ii) the energy gap between them changes linearly with time, and iii)
the diabatic coupling between the two states is constant. In this framework,
ISC is enhanced when the energy gap changes slowly and the diabatic coupling
is large.

\begin{figure}[hbtp]
  \includegraphics[width=\linewidth]{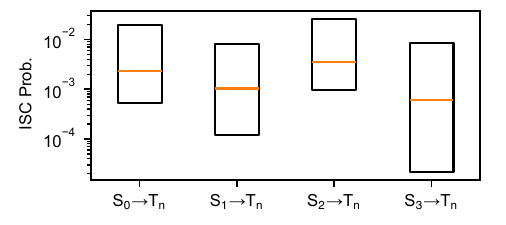}
  \caption{\label{fig:lz_confidence}
    Confidence intervals for the probability of intersystem crossing (ISC) in cyclobutanone.
    The orange line shows the ISC probability using root mean square (RMS) values for
    energy gap rates and spin-orbit couplings. The box shows the range
    of ISC probabilities using energy gap rates and spin-orbit couplings
    that are one standard deviation away from the RMS.
    Maximum values for ISC probabilities for each singlet state to the triplet manifold
    are displayed.
  }
\end{figure}

We estimated the probability of ISC in cyclobutanone by
extracting energy gap rates and spin-orbit coupling matrix elements from
a subset of 9 PBE0-FSSH trajectories at an interval of 0.1 ps.
From these snapshots, we determined the root-mean-square
(RMS) and standard deviation of the energy gap rates and spin-orbit coupling
matrix elements (see Tables SIII - V of the supplementary information).
Finally, we used these to estimate confidence intervals
for the probability of ISC. The results are shown in Figure \ref{fig:lz_confidence}.
From here we see that the probability of ISC is low for all estimates. For example,
using the RMS values, we find the ISC probability to be less than 0.5\% in all cases.
Estimating the ISC probability using energy gap rates that are one standard deviation
lower than the RMS and spin-orbit couplings that are one standard deviation
larger than the RMS, we still find an ISC probability of less than
3\% in all cases.
We emphasize that these are estimating the probability of an ISC
event \emph{each time two states cross in energy}. The actual
ISC probabilities may be lower because not all states cross.
Therefore, we conclude that the overall yield of ISC is likely to be less than 10-20\%
in cyclobutanone, and thus our simulations describe at least 80-90\% of the
photoproduct channels.
We note that this contrasts sharply with the near 100\% yield of ISC
observed in the gas phase photochemistry of
cyclobutanone when excited using 300-320 nm light.\cite{Lee1971JACS2,Lee1971JACS}
We speculate that this discrepancy is due to the higher photon energy used here,
which leads to more nuclear kinetic energy during the photochemical reaction and
thus lower ISC yields within the Landau-Zener picture.
Such a phenomenon has been observed in other photochemical reactions.
For example, the relative rate of internal conversion in hexafluoroacetone
has been found to increase with increasing photon energy.\cite{Knecht1975CPL}

\subsection{Is the ground state ever a diradical?}

We observed from our simulations that the overall photoproduct yield of about
9\% within the first 10 ps of the simulation was lower than expected
based on prior experimental results.
For example, photoproduct yields of 10-22\% are observed in
the liquid phase and estimated to be 30-40\% in the gas
phase when exciting in the 300 nm to 320 nm wavelength
window.\cite{Lee1971JACS2,Lee1971JACS}
Our naive expectation was that the higher photon energy used
here would in turn lead to larger photoproduct yields.
Moreover, prior experimental results found nearly equal C3:C2 ratios,
ranging from about C3:C2 of 1:2 to 2:1,\cite{Lee1971JACS}
whereas our results indicate a C3:C2 ratio of 1:8.
Thus, we test
whether our methodology was systematically underestimating the yield of
photoproducts in general, and the C3 products in particular.

The likely culprit for an anomalously low photoproduct yield would be
use of a restricted Kohn-Sham (RKS) reference for the ground state.
Consequently, diradical or open-shell ground states are
not accessible. Open-shell singlet excited states, on the other hand, are correctly
described within TDDFT as singlet excitations out of the closed-shell reference.
Although all the photoproducts of cyclobutanone are expected to be closed-shell,
we considered the possibility that there could be reaction intermediates with
diradical character. To test this, we performed a series of broken-symmetry
TDDFT simulations to determine whether the ground state of cyclobutanone
becomes diradical.

\begin{figure}[htbp]
  \includegraphics[width=\linewidth]{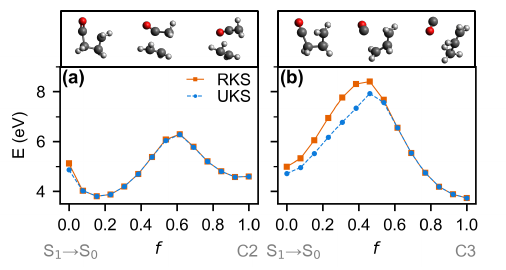}
  \caption{\label{fig:rxn_path}
    Comparison of RKS and UKS energies along representative dissociation
    paths leading to a) C2 and b) C3 photoproducts.
    The reaction path was generated by interpolating between a representative
    hopping structure ($f=0$) and dissociated products ($f=1$) using the linear synchronous
    transit method.\cite{Halgren1977CPL} Energies are relative to the
    \ce{S0} minimum.
    Above each plot are the starting, highest energy intermediate, and ending structures of
    the reaction path.
  }
\end{figure}

First, we wanted to determine whether diradical intermediates might be important
for the formation of photoproducts \emph{during the dynamical simulations}.
To test this, we scanned the potential energy connecting
representative \ce{S1}$\rightarrow$\ce{S0}
hopping structures to either the C2 or C3 photoproducts.
We used the linear synchronous transit\cite{Halgren1977CPL} method to
interpolate between
hopping structures and photoproducts and then computed ground state energies using both
restricted (RKS) and unrestricted Kohn-Sham (UKS) calculations.
We use the highest energy
found for each interpolated path as a rough approximation for the reaction
barrier, although we note that they are in fact upper bounds.
As shown in Figure \ref{fig:rxn_path},
we found no differences between the C2 product reaction barrier computed using
RKS and UKS. On the other hand, broken symmetry solutions were indeed
found along the C3 reaction path. The approximate barrier height
along the C3 path was lower by about 0.5 eV using UKS than was found with RKS.
However, we argue here that this is a \emph{quantitative} rather than \emph{qualitative}
error, because using a UKS reference changes the barrier heights but not the
overall shape of the reaction path. Thus, we conclude that RKS likely
underestimates the yield of C3 photoproducts, but that the qualitative
behavior of photoproduct formation is likely consistent between RKS and UKS.

Next, we wanted to determine whether the photon energy by itself was
sufficient to form photoproducts, and whether a diradical ground state was
essential to that reactivity. To do this, we performed a series of 5 ps
BRS-TDDFT simulations
with initial nuclear velocities sampled from a temperature of
2716 K, 4363 K, and 10000K,
which give total kinetic energies equivalent to the energy of a
387 nm (3.2 eV), 243 nm (5.1 eV) photon, and 87 nm (14.3eV) respectively.
This is equivalent to assuming the photon energy is immediately converted into
thermal energy, and thus we consider the products formed in this case as
thermal products.
For the simulations at 2716 K (3.2 eV) and 4363 K (5.1 eV),
we saw no evidence of thermal product formation
and no evidence of a diradical ground state.
With an initial temperature of 10000 K (14.3eV),
the molecule quickly dissociated, leading to convergence failures for the
BRS-TDDFT calculations. 
These results are at least consistent with our findings from the reaction path analysis,
as the energy difference between the highest energy intermediate
and the optimized ground state minimum is 6.3 eV for C2 dissociation 
and 7.9 eV for C3 dissociation using UKS.
Thus, we conclude that thermal energy of 5.1 eV
is not sufficient to form products on a 5 ps timescale.

%%%%%%%%%%%%%%%%%%%%%%%%%%%%%%%%%%%%%%%%%%%%%%%%%%%%%%%%%%%%%%%%%%%%%%%%%%%%%%%%%%%%%%%%%%

\section{Conclusion} \label{sec:conclusion}

The ultrafast photochemistry of cyclobutanone is rich and complex,
and thus an excellent target for a double-blind benchmark of modern
simulation and experimental techniques.
In this work, we took a pragmatic ``workhorse'' approach to
simulating the photochemistry of cyclobutanone, using TDDFT-based
surface hopping to simulate the excited state dynamics and
ultrafast electron diffraction (UED) patterns. Our simulations indicate that
when excited using 200 nm light, cyclobutanone photoexcites into the \ce{S2}
state, which then decays sequentially from \ce{S2}$\rightarrow$\ce{S1}
with a time constant of about 7.0 ps, and then from \ce{S1}$\rightarrow$\ce{S0}
with a time constant of about 550 fs.
However, the picture is slightly different when considering the
ultrafast electron diffraction patterns. A global and targeted analysis
of the time-resolved $\Delta$PDF indicate two important time scales after the initial
rise of the excited state:
i) a 0.462 ps decay with an associated UED pattern that closely resembles
the $\Delta$PDF of the \ce{S2} minimum geometry, and ii) a longer 16.8 ps
decay that resembles a combination of the \ce{S1} minimum geometry and the
fully separated C3/C2 products.
The difference between the timescales obtained from the electronic states
and the analysis of the time-resolved UED underscores the need for direct
comparisons between experimental and computational observables.

To complement our photodynamics simulations, we performed a series of
sensitivity analyses to estimate the magnitude of the most likely sources
of error. We consider this analysis to be part of a ``pre-mortem'' of our
results, and we use them to answer the question ``If our results end up
with significant qualitative failures, what are the most likely sources
of error?'' In our estimation, the most likely reasons for qualitative failure are
the neglect of intersystem crossing (ISC) and underestimation of the C3 photoproduct
yields.
The restriction to singlet states in our simulations is a technical
limitation of the current implementation which will be addressed in future work.
We estimated overall triplet yields to be less than 10-20\% in cyclobutanone,
but the high degree of uncertainty in this estimate limits our confidence.

We expect our simulations to underestimate the overall photoproduct yield
for two reasons. First, the inability of TDDFT to recover true conical
intersections between the ground state and the first excited state
can overestimate internal conversion pathways and thus miss
some relevant photodecay pathways.\cite{Papineau2024JPCLa}
Second, we found evidence that the dissociation pathway leading to C3
photoproducts likely involves a diradical ground state, which is not
captured with our restricted Kohn-Sham reference. Thus, we expect that
the overall photoproduct yield is likely higher than the 9\% we observed,
and that the C3:C2 ratio is likely closer to 1:1 than the 1:8 we observed.
In addition, even if the overall yield is approximately correct, the
low number of reactive trajectories would lead to high uncertainty in the
C3:C2 ratio. More systematic approaches to simulating surface hopping
swarms would be needed to determine the C3:C2 ratio with higher
confidence.\cite{Parker2020JCPa}

Of course, an additional potential outcome is that the
reason for any discrepancy between our simulations and the
experimental results is completely unforeseen.
Regardless of the outcome of the prediction challenge, we
believe this combination of prediction and uncertainty
estimation will be a valuable approach for photochemistry simulation,
where the reactivities are especially complex.

%%%%%%%%%%%%%%%%%%%%%%%%%%%%%%%%%%%%%%%%%%%%%%%%%%%%%%%%%%%%%%%%%%%%%%%%%%%%%%%%%%%%%%%%%%
\section*{Supplementary Material}
The supplementary material contains detailed statistics for each of our
PBE0-FSSH trajectories, 
details of the implementation of our
simulated electron diffraction patterns, 
along with $sM(s)$ and $\Delta sM(s)$ plots for selected structures,
and extended data from our Landau-Zener analysis. 

%%%%%%%%%%%%%%%%%%%%%%%%%%%%%%%%%%%%%%%%%%%%%%%%%%%%%%%%%%%%%%%%%%%%%%%%%%%%%%%%%%%%%%%%%%
\begin{acknowledgments}
The authors thank Erqian Mao for partaking in discussions
in the planning stages of this manuscript.
This work was supported by a startup fund from Case Western Reserve University.
This work made use of the High Performance Computing Resource in the Core
Facility for Advanced Research Computing at Case Western Reserve University.
\end{acknowledgments}

%%%%%%%%%%%%%%%%%%%%%%%%%%%%%%%%%%%%%%%%%%%%%%%%%%%%%%%%%%%%%%%%%%%%%%%%%%%%%%%%%%%%%%%%%%

\section*{Data Availability}
The code and data that support the findings of this study
are openly available in ``Ultrafast Photochemistry
and Electron Diffraction for Cyclobutanone in the S2 State'' at
\href{https://osf.io/m8sbq}{https://osf.io/m8sbq}.

%%%%%%%%%%%%%%%%%%%%%%%%%%%%%%%%%%%%%%%%%%%%%%%%%%%%%%%%%%%%%%%%%%%%%%%%%%%%%%%%%%%%%%%%%%

\section*{Author Contributions}
% JCP requires author contributions to adhere to the CRediT taxonomy
% https://publishing.aip.org/resources/researchers/policies-and-ethics/authors/
% https://credit.niso.org/contributor-roles-defined/
% https://credit.niso.org/implementing-credit/
\textbf{Ericka Roy Miller:}
    project administration (lead);
    conceptualization (equal);
    methodology (equal)
    investigation (supporting);
    visualization (equal);
    writing - original draft (lead);
    writing - review and editing (equal).
\textbf{Sean J. Hoehn}:
    conceptualization (equal);
    methodology (equal);
    investigation (lead);
    visualization (equal);
    writing - original draft (equal);
    writing - review and editing (equal).
\textbf{Abhijith Kumar:}
    conceptualization (equal);
    methodology (supporting);
    investigation (supporting);
    visualization (supporting);
    writing - original draft (supporting);
    writing - review and editing (supporting).
\textbf{Dehua Jiang:}
    conceptualization (equal);
    methodology (supporting);
    investigation (supporting);
    visualization (supporting);
    writing - original draft (supporting);
    writing - review and editing (supporting).
\textbf{Shane M. Parker:}
    supervision (lead);
    conceptualization (equal);
    methodology (equal);
    visualization (equal);
    writing - original draft (equal);
    writing - review and editing (lead).

%%%%%%%%%%%%%%%%%%%%%%%%%%%%%%%%%%%%%%%%%%%%%%%%%%%%%%%%%%%%%%%%%%%%%%%%%%%%%%%%%%%%%%%%%%

\bibliography{cyclobutanone}

\end{document}